\documentclass
[lengthestimate,tighten,
amssymb,pra,aps,twocolumn,floatfix,tightenlines,superscriptaddress]{revtex4-1}
\usepackage{graphics}
\usepackage{graphicx,color}
\usepackage{subfigure}
\usepackage{float}
\usepackage[ansinew]{inputenc}
\usepackage{graphicx}
\usepackage{amsmath}
\usepackage{amssymb}
\usepackage{dsfont} 
\usepackage{hyphenat}
\usepackage[percent]{overpic}
\usepackage{xcolor}
\usepackage{appendix}
\usepackage{braket}

\begin{document}

\title{Variational Quantum Unsampling on a Quantum Photonic Processor}
\author{Jacques Carolan}
\email{carolanj@mit.edu}
\affiliation{Research Laboratory of Electronics, Massachusetts Institute of Technology, Cambridge, Massachusetts 02139, USA}
\author{Masoud Mohseni}
\affiliation{Google Quantum AI Laboratory, Venice, California 90291, USA}
\author{Jonathan P. Olson}
\affiliation{Zapata Computing Inc., 501 Massachusetts Ave., Cambridge, Massachusetts 02139, USA}
\author{Mihika Prabhu}
\affiliation{Research Laboratory of Electronics, Massachusetts Institute of Technology, Cambridge, Massachusetts 02139, USA}
\author{Changchen Chen}
\affiliation{Research Laboratory of Electronics, Massachusetts Institute of Technology, Cambridge, Massachusetts 02139, USA}
\author{Darius Bunandar}
\affiliation{Research Laboratory of Electronics, Massachusetts Institute of Technology, Cambridge, Massachusetts 02139, USA}
\author{Nicholas C. Harris}
\affiliation{Lightmatter, 61 Chatham St 5th floor, Boston, Massachusetts 02109, USA}
\author{Franco N. C. Wong}
\affiliation{Research Laboratory of Electronics, Massachusetts Institute of Technology, Cambridge, Massachusetts 02139, USA}
\author{Michael Hochberg}
\affiliation{Elenion Technologies, 171 Madison Avenue, Suite 1100, New York, New York 10016, USA}
\author{Seth Lloyd}
\affiliation{Department of Mechanical Engineering, Massachusetts Institute of Technology, Cambridge, Massachusetts 02139, USA}
\author{Dirk Englund}
\affiliation{Research Laboratory of Electronics, Massachusetts Institute of Technology, Cambridge, Massachusetts 02139, USA}
\date{\today}

\begin{abstract}
\noindent
Quantum algorithms for Noisy Intermediate-Scale Quantum (NISQ)
machines have recently emerged as new promising routes towards demonstrating
near-term quantum advantage (or supremacy) over classical systems.
In these systems samples are typically drawn from probability distributions which --- under plausible complexity-theoretic conjectures --- cannot be efficiently generated classically. 
Rather than first define a physical system and then determine computational features of the output state, we ask the converse question: given direct access to the quantum state, what features of the generating system can we efficiently learn?
In this work we introduce the Variational Quantum Unsampling (VQU) protocol, a nonlinear quantum neural network approach for verification and inference of near-term quantum circuits outputs.
In our approach one can variationally train a quantum operation to unravel the action of an unknown unitary on a known input state; essentially learning the inverse of the black-box quantum dynamics.
While the principle of our approach is platform independent, its implementation will depend on the unique architecture of a specific quantum processor.
Here, we experimentally demonstrate the VQU protocol on a quantum photonic processor.
Alongside quantum verification, our protocol has broad applications; including optimal quantum
measurement and tomography, quantum sensing and imaging, and ansatz validation.

\end{abstract}
\maketitle

\section{Introduction} 
\label{sec:introduction}
The construction of a universal error-corrected quantum computer would enable an exponential advantage over the best classical computer in a variety of computational tasks \cite{Nielsen:2011vx, Montanaro:2016iz}.
While significant progress has been made in reducing errors on physical qubits beyond the required fault tolerance levels \cite{Benhelm:2008ki, Barends:2014fu, Gaebler:2016ge}, scaling these systems up to a level required for large-scale computing is a major outstanding challenge \cite{Fowler:2012fi}.
Given this difficulty there has emerged a significant effort towards algorithms for 
NISQ processors, that can solve problems without the need for full-scale error correction \cite{Mohseni17,Preskill2018quantumcomputingin}.
Not only would such a machine reveal a fundamental
gap between the computational power of the quantum and classical
worlds \cite{Boixo:2016un}, they could potentially advance fields such as combinatorics,
\cite{Farhi:2014wk}, quantum simulation \cite{AspuruGuzik:2012ho,
  McClean:2016bs, kokail2018self}, and neural networks \cite{Romero:2017ci, Farhi:2018wv, schuld2018circuit, Chen:2018wx, steinbrecher2018quantum}.

Hardware specific quantum algorithms have been developed to demonstrate a quantum advantage \cite{Aaronson:2011tja, Bremner:2015tb, Fefferman:2015vw, Bouland:2018ue}. 
Moreover, the systems requirements (such as noise or qubit number) to show an unambiguous advantage have been analyzed \cite{Boixo:2016un, Neville:2017do}.
Generally, quantum advantage algorithms for NISQ processors follow a
similar structure: showing that under reasonable complexity-theoretic
conjectures, efficient classical sampling from a distribution $p_U(x)\equiv
|\braket{x|\psi_\text{out}}|^2$ is intractable \cite{Aaronson:2011tja,Boixo:2016un}.
Here $\ket{\psi_\text{out}}=\hat{U}\ket{\psi_\text{in}}$ is a quantum
state generated by a quantum circuit $\hat{U}$ acting on an input state $\ket{\psi_\text{in}}$, and $\{\ket{x}\}$, for example, is the set of bit strings in the computational basis.
As experiments reach the regime where they can no longer be
classically simulated \cite{Bohnet:2016gs,Wang:2016dk, Zhang:2017et,
  Bernien:2017bp, Kelly:2018th} the question of verification becomes
paramount \cite{harrow2017quantum}.
Unlike problems such as factoring which is in the complexity class \textsc{NP} and therefore can be efficiently verified \cite{Shor:1999ul}, sampling problems typically exist outside of this class and efficient verification may not be possible \cite{hangleiter2018sample}.
Machine-level verification techniques have been developed using information about the physical system to achieve efficient verification \cite{Carolan:2014koa, Spagnolo:2014kh}, but a hardware independent approach to verification is outstanding.

\begin{figure}[t!]
\includegraphics[trim=0 0 0 0, clip, width=0.9\linewidth]{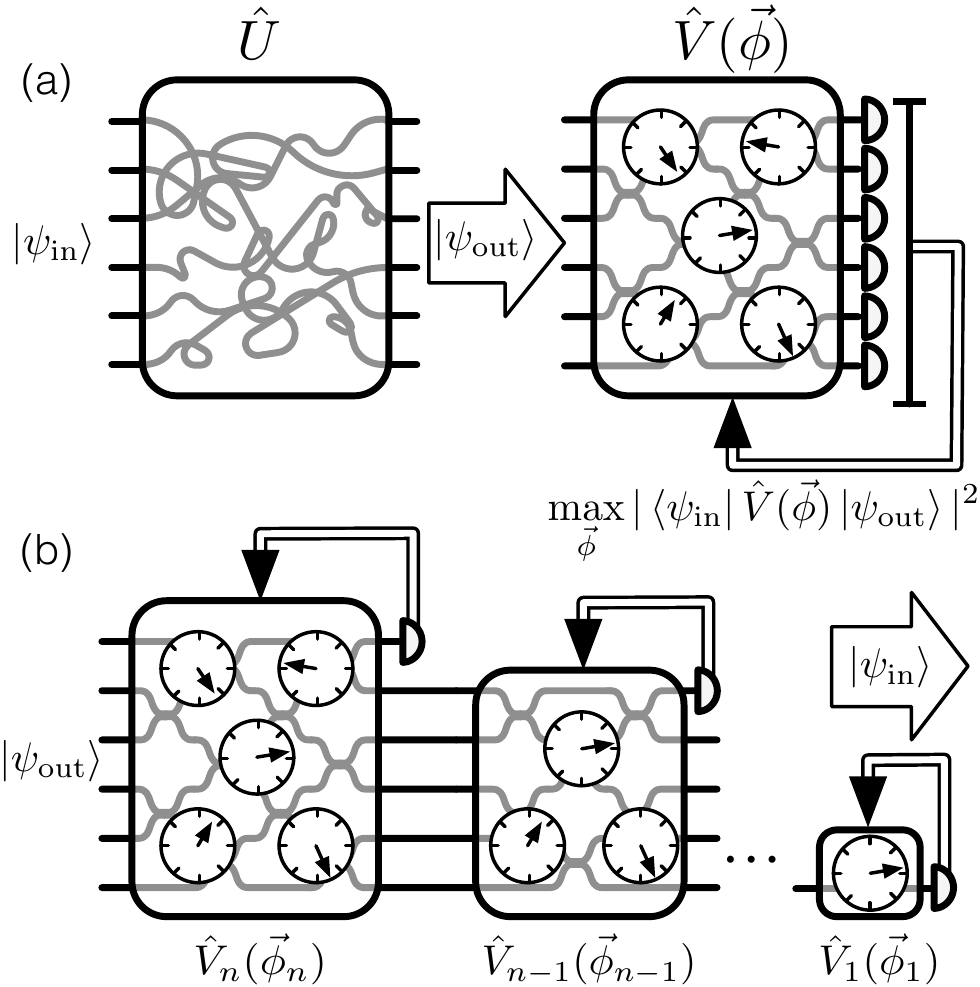}
\caption{\textbf{Variational Quantum Unsampling.}
(a) Given a state $\ket{\psi_\text{out}}=\hat{U}\ket{\psi_\text{in}}$, the task is to find the circuit that returns $\ket{\psi_\text{in}}$, thus determining some features of $\hat{U}$.
The variational quantum unsampling protocol feeds this state into a
controllable quantum circuit and optimizes the parameters $\vec{\phi}$
to find the time reversed condition that
$\hat{V}(\vec{\phi}) \ket{\psi_\text{out}}=\hat{U}^\dagger \ket{\psi_\text{out}}$ over the known input state
(b) Directly optimizing for this condition is inefficient in the qubit number, therefore the layer-wise approach breaks the problem up such that at each stage only a polynomially sized subset of the entire Hilbert space is optimized for.}
\label{fig:fig1}
\end{figure}

Rather than determine properties of an output state given knowledge of the circuit, we ask: given direct access to the state $\ket{\psi_\text{out}}$, can we efficiently learn the physical/computational operation $\hat{U}$, or approximate $\hat{U}$ such that can we generate $\ket{\psi_\text{out}}$?
Here we develop the Variational Quantum Unsampling (VQU) protocol that performs optimization on $\ket{\psi_\text{out}}$ using a controllable auxiliary quantum circuit $\hat{V}(\vec{\phi})$, which is a functional of control parameters $\vec{\phi}$ (see Fig.~\ref{fig:fig1}).
Inspired by neural network approaches to machine learning, our approach approximates the effect of an unknown time-reversed quantum operation $\hat{V}(\vec{\phi}) \approx \hat{U}^\dagger$ to learn the quantum circuit that recovers a \textit{known} input state such that $\hat{V}(\vec{\phi})\ket{\psi_\text{out}} \approx \ket{\psi_\text{in}}$.
In general, our variational learning procedure amounts to partial characterization of unknown unitary operations given knowledge of their actions over certain input states. Consequently, our approach can be understood as a variational approach 
to partial quantum process tomography \cite{Mohseni2008}.
While variational quantum algorithms have been developed for computational tasks such as classification \cite{Farhi:2018wv, schuld2018circuit, mitarai2018quantum, grant2018hierarchical} and simulation \cite{Peruzzo:1kc, Romero:2017ci}, we instead focus on the \emph{verification} of circuit outputs.

\begin{figure*}[t!]
\includegraphics[trim=0 0 0 0, clip, width=0.90\linewidth]{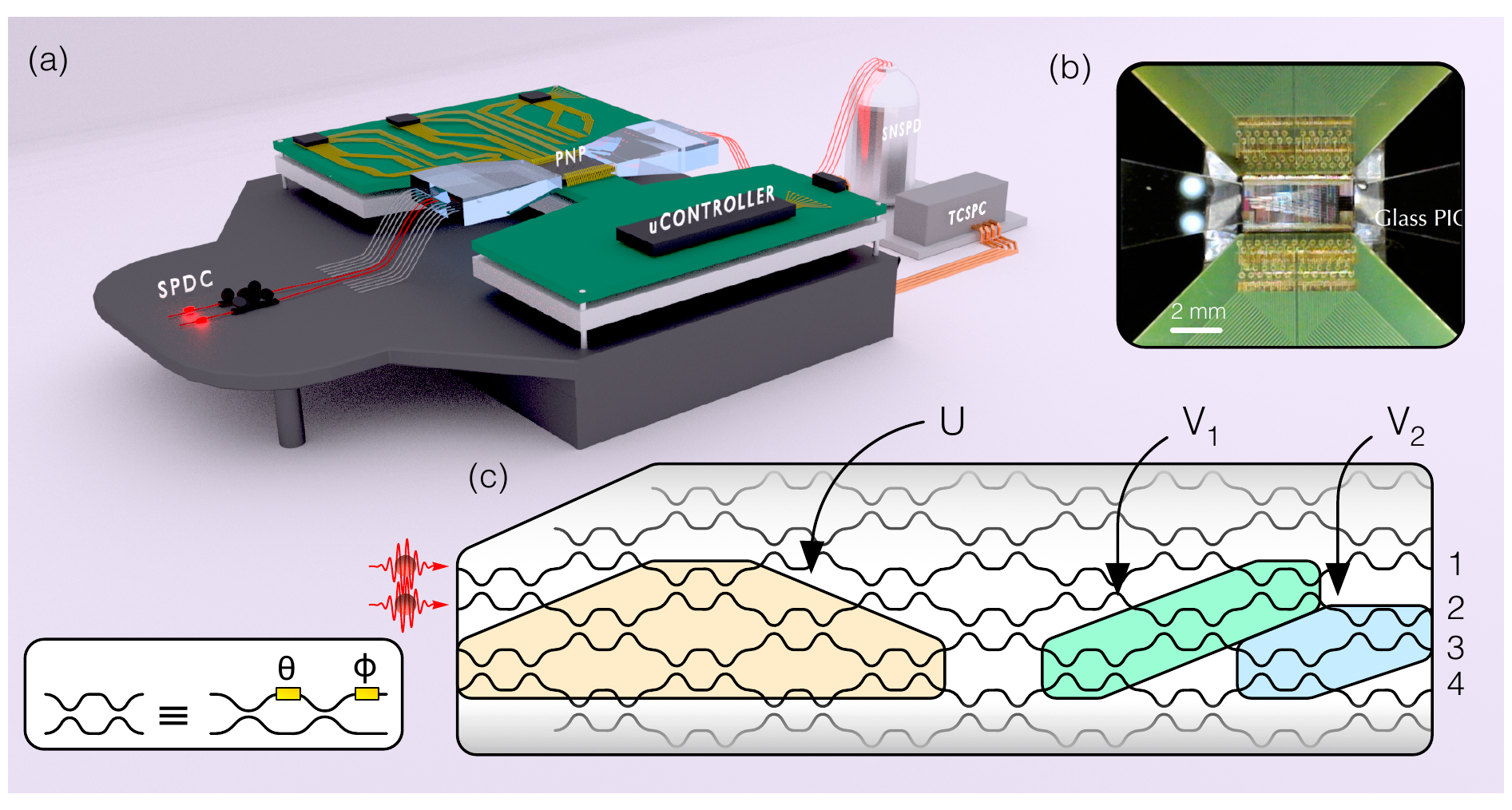}
\caption{\textbf{Optical VQU in a Quantum Photonic Processor.}
(a) Pairs of photons are first generated via spontaneous parametric down-conversion (SPDC) and delivered to a programmable nanophotonic processor (PNP) which is mounted on top of a Peltier cooling system to maintain thermal stability.
After propagating through the circuit photons are out-coupled and delivered to an array of superconducting nanowire single photon detectors (SNSPDs).
Coincidence events are recorded by a time correlated single photon counting system (TCSPC), which are  output to a classical computer which controls a micro-controller unit that drives all 176 on-chip thermo-optic phase shifters.
(b) An optical micrograph of the 26-mode PNP showing all 240 wire-bonds (including grounds) and in/out coupling via two custom built photonic circuits (PIC).  The total footprint of the device is $4.9\times 2.2~\text{mm}$ .
(c) A schematic of the PNP with separate regions marked for the unsampling protocol. 
Each Mach-Zehnder interferometer (MZI) comprises an internal $\theta$ and external $\phi$ phase shifter (inset).
The orange circuit implements the sampling operation, and the green and blue circuits implement the first and second layer of unsampling protocol respectively.
The full protocol requires active control of 46 thermo-optic phase shifters. 
}
\label{fig:fig2}
\end{figure*}

The principle of our approach is architecture-independent, however quantum optics provides a natural platform to explore VQU protocols for two primary reasons:
First, a clear and well defined near-term quantum algorithm native to quantum optics has been developed known as boson sampling \cite{Aaronson:2011tja, Olson_2018}.
Second, advances in integrated quantum photonics \cite{Politi:2008tl} have enabled the demonstration of large-scale reconfigurable quantum circuits that can implement any unitary operation across optical modes \cite{Carolan:2015vga, Harris:18}.
State-of-the-art  `universal linear optical processors' now contain $\sim 200$ control parameters across $\sim 30$ spatial modes \cite{Harris:2017hi}. 

In this letter we first introduce the unsampling problem and show that the major challenge lies in selecting efficiently accessible subsets of the entire Hilbert space for optimization.
We then propose the VQU protocol, a general layer-by-layer training approach which has been shown to successfully train classes of deep-neural networks that otherwise get stuck in local optima \cite{Hinton:2006dk, Bengio:vb, Brock:2017ws,Hettinger:2017wg}.
We develop an optical VQU algorithm that sequentially unsamples the boson sampler mode-by-mode
and perform a proof-of-concept demonstration of this protocol with a quantum photonic processor (QPP).
Altogether our results point towards a new practical approach to quantum verification, which will not only find application in quantum optical systems, but general NISQ processors as they push limits of classical computation.

\section{Variational Learning} 
\label{sec:formalisation_of_the_problem}
Formally the unsampling problem asks: given direct access to a polynomial number of copies of $\ket{\psi_\text{out}}$, find a circuit that returns the known input state $\ket{\psi_\text{in}}$, thus determining some elements of $\hat{U}$.
Prima facie, one can imagine taking $\ket{\psi_\text{out}}$ and coherently passing it through an appropriately parametrized circuit $\hat{V}(\vec{\phi})$ [see Fig.~\ref{fig:fig1}(a)].
In the language of machine learning, we can define a loss function 
\begin{equation}
\label{eq:ineff}
	L(\vec{\phi}) = 1 - |\braket{ \psi_\text{in} | \hat{V}(\vec{\phi}) |\psi_\text{out}} |^2
\end{equation}
that quantifies the distance between the output state and the input state, and is bounded $L(\vec{\phi})\in[0,1]$.
Searching for the condition that
\begin{equation}
	\min_{ \vec{\phi} } L(\vec{\phi}) = 0
\end{equation}
leads to $\hat{V}(\vec{\phi}) \approx \hat{U}^\dagger$ over a given input
state. That is, the circuit which generates $\ket{ \psi_\text{in}}$ is found, corresponding to a single column of $\hat{U}$.  
Note however that without a well chosen ansatz we could have $|\braket{ \psi_\text{in} | \hat{V}(\vec{\phi}) |\psi_\text{in}} |^2\approx 1/D$ where $D$ is the dimension of the system, which typically scales exponentially in the particle number.
The probability for an individual event is therefore exponentially unlikely and estimating the associated probability takes exponential time.
Moreover, it has recently been shown that gradient-based quantum circuit learning becomes exponentially inefficient due to a very flat loss
landscape if one starts with a generic random initial state over the entire Hilbert space \cite{McClean2018}, however initialization strategies have been proposed to overcome this obstacle \cite{grant2019initialization}.  
Here we opt for a divide-and-conquer approach that selects efficiently accessible subspaces of the entire Hilbert space for the stochastic optimization, within a layer-wise model of learning. 

Towards this end we use multiple unitaries $\hat{V}_k$, or `layers', and a layer-by-layer training approach, that at each stage optimizes over only a polynomially sized subset of the full Hilbert space [see Fig.~\ref{fig:fig1}(b)].
To illustrate this procedure consider a $n$-qubit system with the known pure initialization state, with tensor product structure (such as a mean-field state), $\ket{\psi_\text{in}}=\ket{\alpha_1,\alpha_2 \dots \alpha_n}$, where $\ket{\alpha_i}$ is the state of the $i^{\text{th}}$ qubit.
The first training stage feeds $\ket{ \psi_\text{out}}$ into a circuit $\hat{V}_n(\vec{\phi}_n)$ acting on all $n$ qubits.
Letting $\rho_1=$ $\hat{V}_n(\vec{\phi}_n) \ket{\psi_\text{out}} \bra{\psi_\text{out}} \hat{V}^\dagger_n(\vec{\phi}_n)$, the optimization then varies circuit parameters $\vec{\phi}_n$ to minimize
\begin{equation}
\label{eq:loss}
	L_1(\vec{\phi}_n) = 1 - \braket{ \alpha_1 | \text{Tr}_{2\dots n} (\rho_1) |\alpha_1}.
\end{equation}
If $L_1(\vec{\phi}_n) = 0$ then the first qubit is successfully found in the state $\ket{\alpha_1}$ and the remainder of the qubits $\rho'_1=\text{Tr}_1(\rho_1)$ are in a pure state.
The state $\rho'_1$ is then fed into a circuit $\hat{V}_{n-1}(\vec{\phi}_\text{n-1})$ acting on the remaining $n-1$ qubits and $L_2(\vec{\phi}_\text{n-1})$ is minimized to maximize the overlap between the second qubit and $\ket{\alpha_2}$.
This process is repeated for $n$ stages and if successful each subsequent stage will disentangle a qubit until the total output is $\ket{\alpha_1,\alpha_2 \dots \alpha_n}$.
Critically, the probability estimated at each stage is now exponentially boosted, with $L_i(\vec{\phi})$ scaling as $\mathcal{O}(1)$ (independently of $n$).
Moreover, the error in a single stage of unsampling $L(\vec{\phi})\approx \epsilon$ should scale as $\epsilon \ll 1/n$, such that as $n$ becomes large the overall unsampling fidelity does not vanish.

Our protocol enables a two-fold approach to verification.
First, the solution unitary is given by $\hat{V}_\text{sol}=\prod^{n}_{i=1} \hat{I}_{n-i} \otimes \hat{V}_i(\vec{\phi}_i)$ (where $\hat{I}_j$ is the identity operation acting on the first $j$ qubits) which enables direct verification of the sampling circuit. 
Second, deviations from $\ket{\psi_\text{in}}$ signals decoherent error in the sampling protocol, which can further be inspected by tomography on a reduced subset of qubits.
A layer-wise training approach with conditional feedforward was recently used for quantum state discrimination \cite{Chen:2018wx} and recognizing quantum states of matter \cite{cong2018}.

\begin{figure*}[t!]
\includegraphics[trim=0 0 0 0, clip, width=1.0\linewidth]{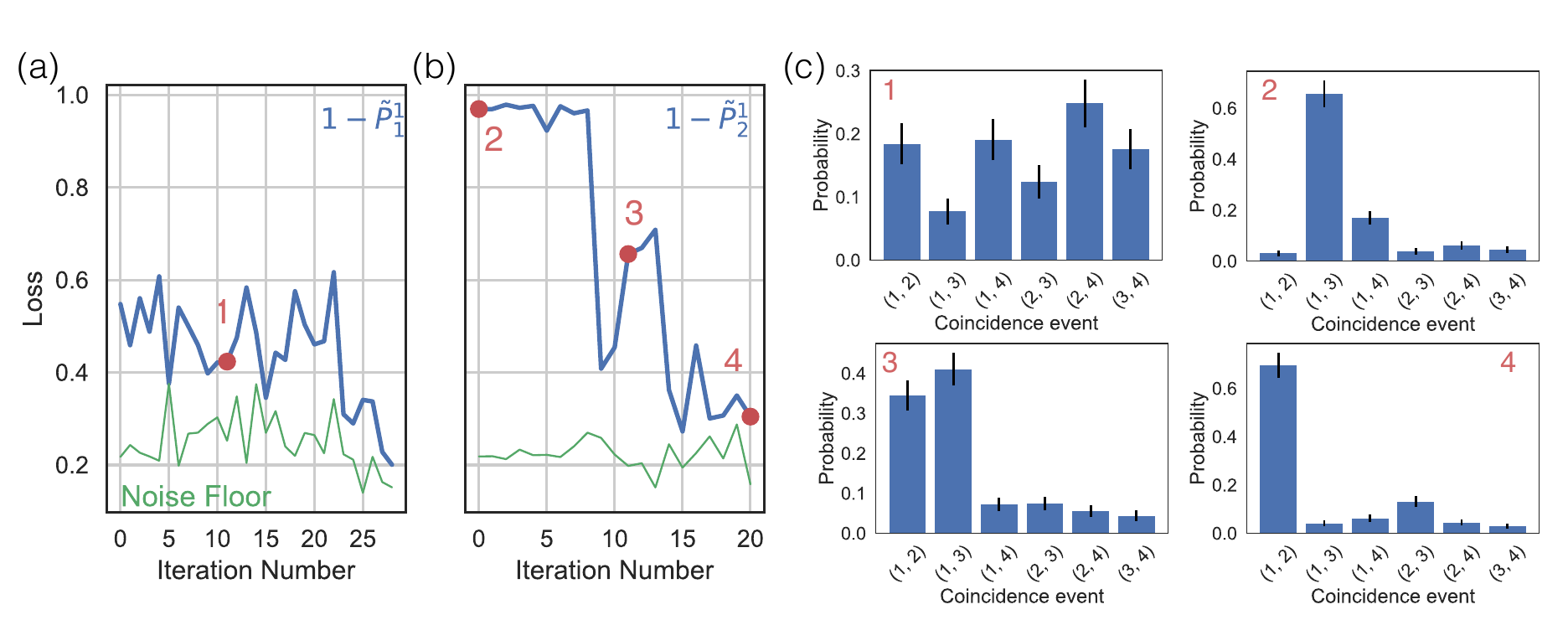}
\caption{\textbf{Experimental Results.}
Variation in the loss function during the optical VQU protocol.
(a) The first layer (blue) minimizes $L_1=1-\tilde{P}^1_1$ to find a photon in the first optical mode.
(b) The second layer (blue) minimizes $L_2=1-\tilde{P}^1_2$ to find a photon in the second optical mode.
The green line plots the experimental noise floor (see text).
Red points correspond to probability distribution time slices as shown in (c).
(c) The probability for all six two-fold coincidence events are plotted with error bars assuming Poissonian counting statistics.
The first plot (top left) shows support across all coincidence events while at the end of the VQU protocol (bottom right) $P=0.695 \pm 0.053$ is found in the $(1,2)$ coincidence event, corresponding to the initialization state $\ket{1_1 1_2}$.
}
\label{fig:fig3}
\end{figure*}

We numerically tested this protocol for up to five qubits.  These simulations, which are provided in Appendices \ref{sec:numerical_methods} and \ref{sec:vqu_qubit_numerics}, converged to numerical precision in all instances.
We conjecture this efficacy is due to an over-parameterization effect:
each layer effectively decouples a qubit from the remainder of the state, and there may be many such circuit settings $\vec{\phi}$ which achieve this condition, whereby $L_i(\vec{\phi})=0$. 
For certain classes of classical deep neural networks, this over parameterization has been shown to both increase expressiveness \cite{Eldan:2015uc} and accelerate training \cite{Arora:2018vn}. 
Similar layer-wise training approaches have found success in training particular classes of classical neural networks without the need for backpropagation \cite{Hinton:2006dk, Bengio:vb, Brock:2017ws,Hettinger:2017wg}.

The above example does not specify how one can physically build the circuit that implements the unitary, as in general constructing an arbitrary unitary requires a circuit depth that grows exponentially in the number of qubits \cite{Barenco:1995hx}.
Each VQU protocol can leverage unique structure in the specific sampling problem to construct a reduced family of unitaries that \emph{can} be efficiently implemented.
If the system parameterization (viz.\ ansatz) $\mathcal{A}_U$ is known, then VQU can be use to characterize the sampling circuit.
However, if the system parameterization is unknown or not known to be optimal, VQU can be used to assess whether a test ansatz $\mathcal{A}_V$ can represent the state, or set of states given by $\mathcal{A}_U$. We refer to this procedure as `ansatz validation', which is related to quantum circuit compiling \cite{heyfron2018, jones2018compilation}.
While it remains an open question exactly which ansatz are amenable to VQU,
in Appendix~\ref{sec:ansatz_validation} we give an example of one such ansatz that is related to the fractional quantum Hall effect. 

\vspace{-.4cm}

\section{Optical VQU}
\label{sec:efficient_subspaces}

Boson sampling is a mathematical proof that shows ensembles of indistinguishable photons when acted on by linear optical circuits (arrays of beamsplitters and phase shifters), generate samples from a probability distribution that cannot be efficiently generated classically \cite{Aaronson:2011tja}.
Formally, given an $n$-photon initialization state of one photon per mode $\ket{\psi_\text{in}}=\ket{1_1 1_2\dots1_n}$ (where $\ket{i_j}$ represents $i$ photons in the $j^\text{th}$ optical mode), each amplitude of the output state $\ket{\psi_\text{out}}=\hat{U}_m \ket{\psi_\text{in}}$ is given by the permanent of a unique $n\times n$ submatrix of the $m$-dimensional unitary $\hat{U}_m$ \cite{Scheel:2004tt}.
The output distribution  $p_{U}(x)=|\braket{x|\psi_\text{out}}|^2$ is therefore also related to permanents, a notoriously difficult function to calculate \cite{Valiant:1979dl}, with $\{\ket{x}\}=\{\ket{i_1 i_2 \dots i_m} \}$ being the set of collision free computational basis states such that $\sum_j i_j = n $ with $i_j\leq 1$.

In optics, an arbitrary $m$-dimensional unitary operator $\hat{U}_m$ across $m$ optical modes can always be constructed out of $m(m-1)/2$ reconfigurable beamsplitters and phase-shifters \cite{Reck:1994dz}.
This theorem therefore provides an efficient circuit ansatz for the optical VQU protocol.
Limiting our discussion to the regime of $n$ photons in $m=n^2$ optical modes, the optical VQU protocol first feeds $\ket{\psi_\text{out}}$ into a $n^2$-dimensional circuit $\hat{V}_1(\vec{\phi}_n)$ and minimizes the loss function \eqref{eq:loss}, which maximizes $\tilde{P}^1_1$, the probability of one and only photon in the first mode.  Note, in the optical case $\ket{\psi_1}=\ket{1_1}$ and the trace operation occurs over the optical mode basis.
Critically, the probability of exactly one photon in the $i^\text{th}$ optical mode $\tilde{P}^1_i$ scales as $\mathcal{O}(1/n)$ and is therefore an efficiently accessible measurement (see Appendix~\ref{ssub:scaling_measurement} for a proof of this).
Given $L_1(\vec{\phi}_n)\approx 0$, the second layer $\hat{V}_2(\vec{\phi}_{n-1})$ is an $n^2-1$ mode circuit acting on $n-1$ photons that maximizes the probability of one photon in the second optical mode.
The optical VQU protocol proceeds for a total of $n$ layers until the initialization state $\ket{1_11_2\dots1_n}$ is recovered.

Each layer contains at most $\mathcal{O}(n^4)$ parameters however careful analysis of the circuit structure reveals that $\mathcal{O}(n^2)$ parameters is sufficient for each layer (see Appendix~\ref{sub:circuit_construction}), so the full protocol requires $\mathcal{O}(n^3)$ parameters.
Once again, in certain cases the over-parameterization may in fact accelerate optimization.
In Appendix~\ref{sec:squeezing_photons_into_modes} we also give an alternate unsampling protocol that first compresses $n$ photons into the first $n$ modes, and may be more practical for implementation.

\section{Experimental unsampling} 
\label{sec:Experimental_Unsampling}

We implement a proof-of-concept demonstration of the optical VQU procedure on a state-of-the-art 
quantum photonic processor comprising three stages: (1) photon generation, (2) reprogrammable quantum circuitry and (3) single photon detection, all within an actively configured feedback loop for optimization [see Fig~\ref{fig:fig2}(a)].

Pairs of degenerate photons at 1582~nm are generated via spontaneous parametric down-conversion (SPDC) from a custom-fabricated periodically-poled KTiOPO4 (PPKTP) crystal under extended phase-matching functions \cite{Chen:2017gt}.
Photon pairs are then collected into optical fibers and delivered to a programable nanophotonic processor (PNP) \cite{Harris:2017hi} via a custom built optical interposer which reduces the mode field diameter of the input fibers to better match that of silicon waveguides [Fig~\ref{fig:fig2}(b)]. 
The PNP consists of 176 individually tuneable phase shifters across 26 optical modes, fabricated in a CMOS compatible silicon photonics process.
On-chip Mach-Zehnder interferometers (MZIs) are controlled via two phase shifters, with an internal phase shift $\theta$ for splitting ratio configuration and external phase shift $\phi$ for phase configuration.
A total of 88 MZIs are arranged in a mesh that enables different regions of the device to be used for separate quantum operations. In Fig.~\ref{fig:fig2}(c) the sampling circuit is shown in orange, and the unsampling layers are shown in green and blue.

After passing through the PNP photons are out-coupled and delivered to four tungsten silicide superconducting nanowire single photon detectors (SNSPDs) with $\sim 65\%$ quantum efficiency for photon counting.
Correlations across each channels are recorded by a time correlated single photon counting (TCSPC) system which is then fed to a classical computer for processing.
Based on recorded coincidence events across all $\binom{4}{2}=6$ coincidence channels $\{(1,2),(1,3),(1,4),(2,3),(2,4),(3,4)\}$ (where $(i,j)$ represents a coincidence event between optical modes $i$ and $j$), a classical optimizer running the local derivative-free \texttt{BOBYQA} algorithm \cite{Powell:2gDjtIQ0} varies the PNP layer phases to minimize a user defined loss function.

\begin{figure}[t!]
\includegraphics[trim=0 0 0 0, clip, width=0.90\linewidth]{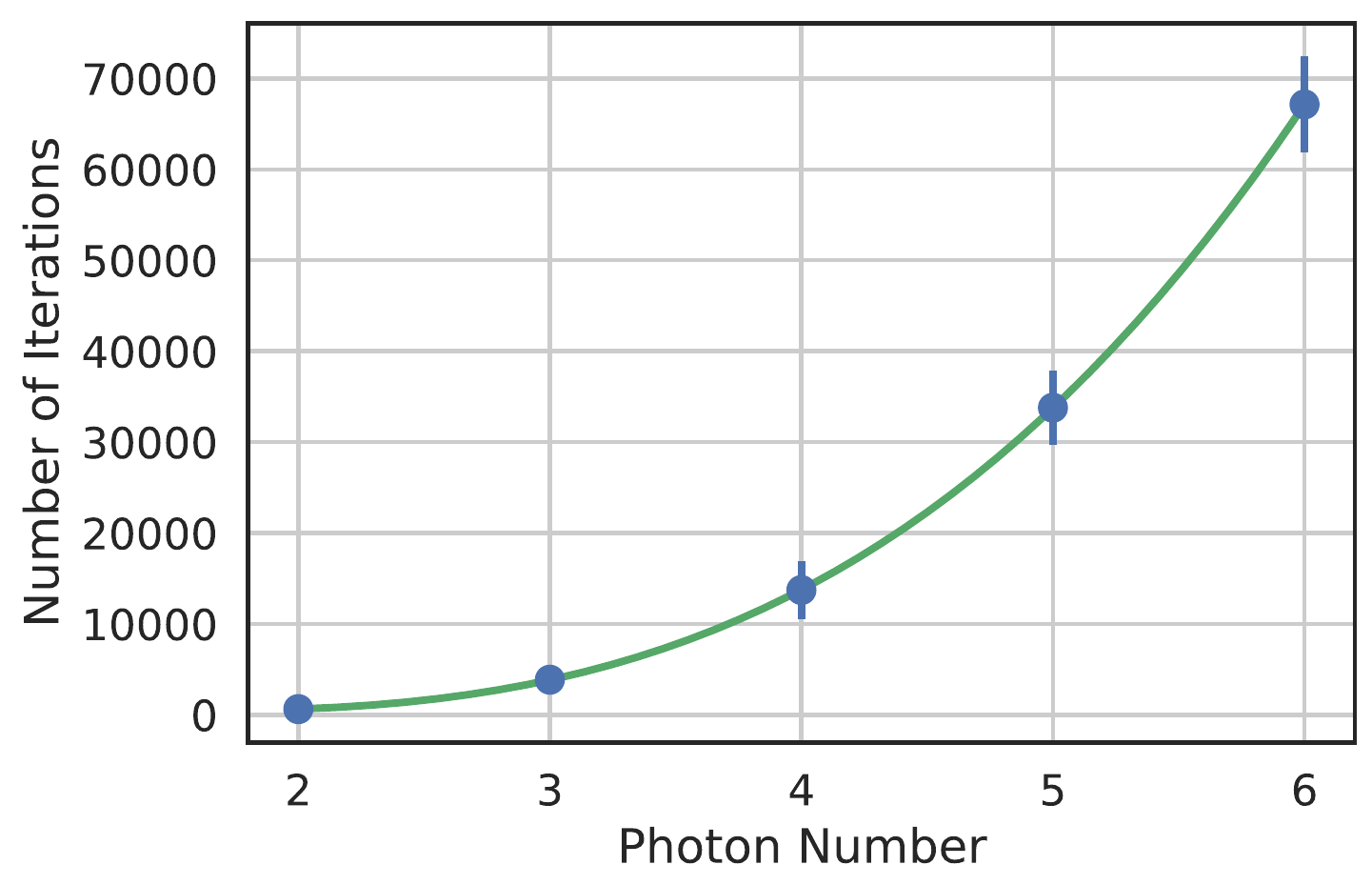}
\caption{\textbf{Monte Carlo Numerics.}
Full boson unsampling Monte Carlo numerics for up to $n=6$ photons in $m=36$ modes.  
The total number of steps required to converge to $\mathcal{F}=1-10^{-5}$ for each photon number is plotted in blue, alongside an expected third order polynomial fit in green with $R^2=1-10^{-6}$.  
Error bars represent one standard deviation from $N=100$ runs.
}
\label{fig:fig4}
\end{figure}

The sampling circuit (Fig.~\ref{fig:fig2}(c), orange) directly dials six MZIs (12 phases) to generate a four-mode random unitary according to the Haar measure \cite{Russell:2017cz}.
Two photons pass through the sampling circuit and the output state is fed into the first unsampling layer (Fig.~\ref{fig:fig2}(c), green); a four-mode circuit acting on modes $\{1,2,3,4\}$.
The classical optimizer is programmed to find a single photon in optical mode 1 by minimizing $L_1(\vec{\phi}_4)=1-\tilde{P}^1_1(\vec{\phi}_4)$.
Each iteration of the optimization collects approximately 100 two-fold coincidence events.
As shown in Fig.~\ref{fig:fig3}(a), the phases $\vec{\phi}_4$ are first randomly initialized and $L_1=0.55$ and after 28 iterations of the optimization $L_1=0.20$, which is at the noise floor of our experiment.
The output state is then fed to the second unsampling layer; a three-mode circuit acting on modes $\{2,3,4\}$ (Fig.~\ref{fig:fig2}(c), blue).
The optimizer is set to find a single photon in mode 2 by minimizing $L_2(\vec{\phi}_2)=1-\tilde{P}^1_2(\vec{\phi}_2)$.  
Crucially, by leaving mode 1 untouched $L_1$ cannot increase.
The phases $\vec{\phi}_2$ are randomly initialized and $L_2=0.97$.  As shown in Fig.~\ref{fig:fig3}(b), after just 20 stages of optimization $L_2=0.31$, which is $1.3\sigma$ from the noise floor of our experiment.

The final fidelity of the VQU protocol, defined as the overlap between the initialization state and output state $\mathcal{F}=|\braket{\psi_\text{in}|\psi_\text{out}}|^2$, was found to be $\mathcal{F}=0.695\pm 0.053$, which is $1.2\sigma$ from the maximal achievable fidelity given the noise floor of our experiment (Fig.~\ref{fig:fig3}(a, b), green).
This noise floor is primarily due to a low signal-to-noise ratio, caused by photon emission from our thermo-optic phase shifters and high fiber-to-chip coupling loss (-8~dB facet to facet).
The deviation from the maximum possible fidelity is likely due to the performance of our optimizer in the presence of finite counts.
Future implementations will use either low loss in/out couplers \cite{Notaros:2016vw} or on-chip single photon sources \cite{Silverstone:2013fu, Carolan:19} and detectors \cite{Najafi:2015ey} to increase the signal-to-noise ratio and boost fidelity.

Alongside the proof-of-concept experimental demonstration extensive numerical simulations were performed for up to six photons.
In Figure~\ref{fig:fig4} we plot the number of iterations required to converge to a fidelity of $\mathcal{F}=1-10^{-5}$, alongside an expected cubic fit for $n=100$ runs (see Appendix~\ref{sec:optical_vqu_numerics} for further details).
The efficiency of these numerical experiments suggest that the presence of local optima are limited and unlikely to prevent convergence for optical unsampling experiments.

\section{Concluding Remarks}
We have introduced the VQU protocol: a nonlinear quantum neural network approach for verification and inference of near-term quantum processors.
Our protocol leverages a divide-and-conquer approach that selects efficiently accessible subspaces of the entire Hilbert space for optimization.
Within a layer-wise learning model, we simulate the effect of an unknown time-reversed quantum operation to recover a known input state.
We demonstrated this protocol optically on a quantum photonic processor.
Our approach can directly be applied to the verification and certification of circuit outputs, and for the comparison and training of circuit ansatz.
Moreover, VQU could also lend itself to the characterization of other physical processes that can be probed by quantum signals such as molecular excitations \cite{brinks2010visualizing}.
Applied to optical systems, VQU may find application as a subroutine in quantum cryptographic protocols \cite{PhysRevX.4.011016}, or for optimal receivers for optical communications \cite{guha2011structured}.
As quantum processors push the limits of what is classically simulable and coherent control of quantum phenomena advances, the problem of quantum state and circuit verification represents a formidable challenge.
We therefore anticipate VQU in particular, and other layer-wise learning models more generally, serving as a vital tool in the arsenal of the quantum engineer.

\begin{acknowledgments}
We thank  E. Farhi, E. Grant, D. Hangleiter, I. Marvian, J. McClean, M. Pant, P. Shadbolt, S. Sim and M. Schuld for insightful discussions.
This work was supported by the AFOSR MURI for Optimal Measurements for Scalable Quantum Technologies (FA9550-14-1-0052) and by the AFOSR program FA9550-16-1-0391, supervised by Gernot Pomrenke. 
J.C. is supported by EU H2020 Marie Sklodowska-Curie grant number 751016.
\end{acknowledgments}

\clearpage

\appendix

\section{Numerical Methods}
\label{sec:numerical_methods}
Whilst the protocols we present in the body of the manuscript are agnostic to the particular optimization algorithm, in practice we necessitate two conditions: (1) the optimization algorithm should be \emph{local} so as to converge efficiently and (2) \emph{gradient free} as in general the gradients will not be a priori known.
We have determined through extensive numerical studies that the \texttt{BOBYQA} algorithm \cite{Powell:2gDjtIQ0} performs well in terms of speed and accuracy, satisfies (1), (2) and is readily implemented in the \texttt{NLOPT} library \cite{Johnson:2011}. 
Consequently all numerical experiments presented in this manuscript use this algorithm.

\section{VQU Qubit Numerical Experiments}
\label{sec:vqu_qubit_numerics}

In the following we present numerical results for the Variational Quantum Unsampling (VQU) protocol performed for up to 5 qubits.  
In general performing an arbitrary operation requires a circuit depth exponential in the number of qubits.
Therefore each implementation of VQU will use a circuit parameterization unique to the problem of interest, which can efficiently implement a family unitaries sufficient to unentangle the qubit.
In these numerics we implement an arbitrary operation parametrized by $2^{2n}-2^{n}$ real numbers $\vec{\phi}_n$ via then encoding of Clements et al., \cite{Clements:2016tv}.
The protocol proceeds as follow:

\begin{enumerate}
	\item Generate the $n$-qubit initialization state $\ket{\psi_\text{in}}=\ket{0}^{\otimes n}$ and pass it through an $n$-qubit sampling operation, chosen randomly according to the Haar measure
	\item Apply an $n$-qubit operation $\hat{V}(\vec{\phi})_n$ and maximize the probability of finding the $j=1$ qubit in the state $\ket{0}_1$
	\item Apply an $(n-1)$-qubit operation $\hat{V}(\vec{\phi})_{n-1}$ spanning qubits $[2,n]$ and maximize the probability of finding the $j=2$ qubit in the state $\ket{0}_2$
	\item Repeat this protocol for $j\in[3,n]$
\end{enumerate}

\begin{figure}[t!]
\includegraphics[trim=0 0 0 0, clip, width=0.9\linewidth]{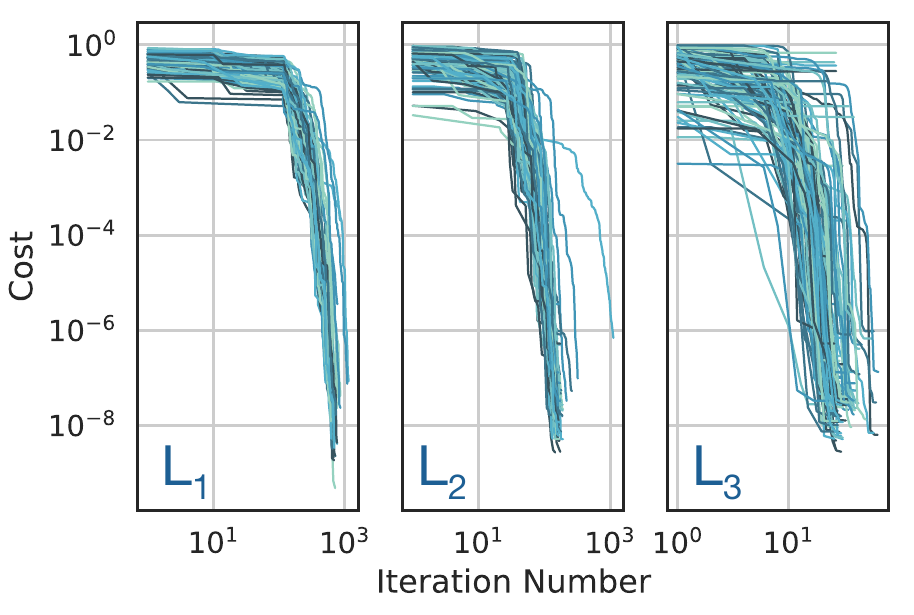}
\caption{\textbf{Three qubit unsampling.} 100 numerical optimizations plotted for the three qubit VQU protocol. 
Each panel plots the results of optimizations to find the $j^{\text{th}}$ qubit in the state $\ket{0}_j$ for $j\in[1,3]$.}
\label{fig:fig1_app}
\end{figure}

\begin{figure}[t!]
\includegraphics[trim=0 0 0 0, clip, width=0.9\linewidth]{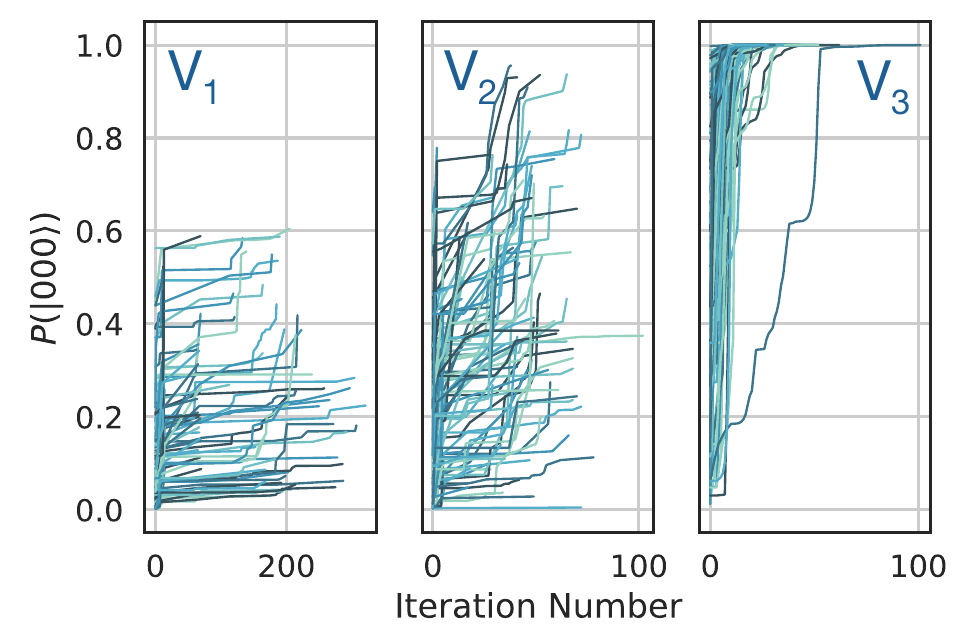}
\caption{\textbf{Initialization state probability.} For the three qubit VQU protocol, we plot the probability of finding the initialization state $\ket{000}$. 
Each panel plots the probability for a different optimization layer $\hat{V}_i$.}
\label{fig:track000}
\end{figure}

We perform 100 Monte Carlo numerical experiments for up to five qubits.
We allow for random restarts in case a desired threshold of $1\times 10^{-5}$ is not achieved. 
In all cases numerical accuracy is achieved.
Note, that sampling operations are chosen according to the Haar measure so as to avoid bias in our numerical experiments.  However it is likely such circuits would suffer from the `barren plateaus' problem \cite{McClean2018}, whereby the gradient becomes exponentially small with the number of qubits.
Initialization strategies have been proposed to overcome this limit \cite{grant2019initialization}, which are also compatible with VQU.
Indeed, the layer-wise training approach we present may enable a reduction of circuit depth, potentially mitigating the effects of barren plateaus.
In Fig.~\ref{fig:fig1_app} we plot each optimization run for the three qubit case.
As the protocol progresses, the optimizing circuit acts on fewer qubits and thus requires less parameters. Consequently the number of iterations required to reach numerical accuracy becomes fewer.
In Fig.\ref{fig:track000}, for the three qubit case, we track the probability of finding the initialization state $\ket{\psi_\text{in}}=\ket{000}$.  Note that while this quantity is never directly optimized, it does reach unity as required.

\section{Ansatz Validation}
\label{sec:ansatz_validation}
In the following we formalize the notion of ansatz validation described in the main text.
The ability of a shallow depth quantum circuit to implement a specific operation (or set of operations) relies on a well chosen ansatz (i.e. circuit parametrization).
If the ansatz is known VQU can be used to characterize the sampling circuit.  
If however the ansatz is unknown, or not known to be optimal VQU can be used to \emph{validate} a test ansatz.
Formally, given a system ansatz $\mathcal{A}_U$ that generates either a single instance of a circuit/state or set of circuits/states, ansatz validation tests whether a trail ansatz $\mathcal{A}_V$ is capable of representing the system features and thus $\mathcal{A}_U \in \mathcal{A}_V$.
If $\mathcal{A}_V$ requires fewer gates than $\mathcal{A}_U$, but can reproduce many of the salient features, then this can be seen as an instance of circuit compilation.
To perform ansatz validation we parameterize the unsampling circuit $\hat{V}$ with a trail ansatz $\mathcal{A}_V$ and perform VQU.  The fidelity of the full protocol thus quantifies how well the trail ansatz can represent the system.

\begin{figure}[t!]
\includegraphics[trim=0 0 0 0, clip, width=0.9\linewidth]{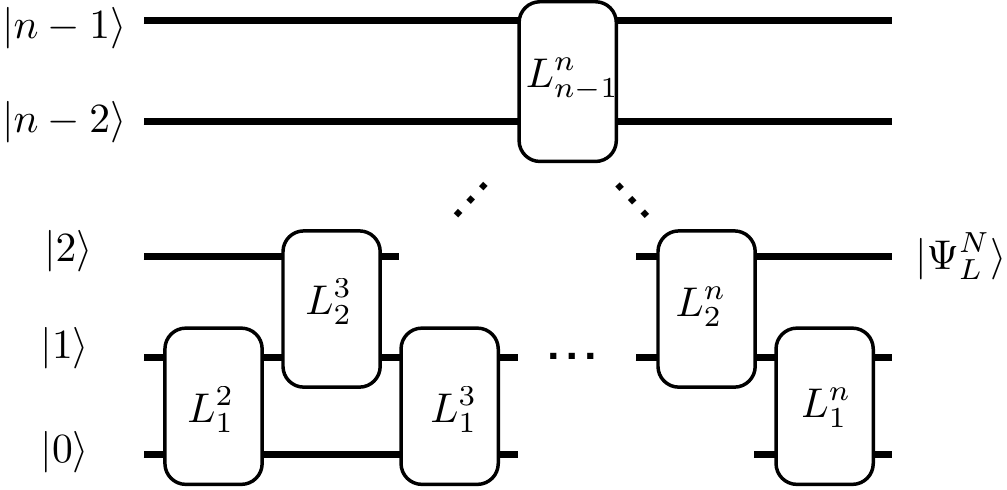}
\caption{\textbf{Laughlin states.}  The ansatz to generate $n$-qudit Laughlin states with filling fraction one.}
\label{fig:laughlin}
\end{figure}

As an example of an ansatz which can be validated by VQU, we examine a proposal by Latorre et al. \cite{PhysRevA.81.060309} for a family of quantum circuits that can generate Laughlin wave functions, which are conjectured to be ground-states of the fractional quantum Hall effect \cite{PhysRevLett.48.1559, PhysRevLett.50.1395}.
In this work they construct an ansatz for a system of $n$ qudits that can generate the Laughlin states $\ket{\Psi_L^{n}} $ (written in terms of single particle angular momentum eigenstates) with a filling fraction of one.  Here
\begin{equation}
	\ket{\Psi_L^{n}} = \frac{1}{n!}\sum_\mathcal{P} \text{sgn}(\mathcal{P}) \ket{a_1, \dots, a_n}
\end{equation}
where $\mathcal{P}$ is the set of all possible $n!$ permutations of the set $\{0,1,\dots,n-1\}$, and the relative sign of the permutation corresponds to the parity of the number of transpositions required to transform one state into the other.
The general circuit to construct an $n$-qudit state is shown in Fig.~\ref{fig:laughlin}, where $L^{n+1}_k= \prod_{i=1}^{n-1} W_{i,n}(1/(k+1))$ and $W_{i,j}(p)$ is the two qudit operation
\begin{equation}
	\begin{aligned}
	W_{i,j}(p) \ket{ij} &= \sqrt{p} \ket{ij} -\sqrt{1-p} \ket{ji} \\
	W_{i,j}(p) \ket{ji} &= \sqrt{1-p} \ket{ij} +\sqrt{p} \ket{ji},
	\end{aligned}
\end{equation} 
with $W_{k,l}\ket{ij}=\ket{ij}$ if $(k,l) \neq (i,j)$.  The input state is the product state $\ket{n-1, n-2, \dots 0}$ with the full circuit requiring $\mathcal{O}(n^2)$ gates and depth $\mathcal{O}(n)$, thus scaling efficiently.

To unsample Laughlin states we construct an $n$-qudit auxiliary quantum circuit $\hat{V}(\vec{\theta})^n_L$, which comprises $\tilde{W}$, the conjugate transpose of the $W$ operator
\begin{equation}
	\begin{aligned}
	\tilde{W}_{i,j}(p) \ket{ij} &= \cos(\theta) \ket{ij} + \sin(\theta) \ket{ji} \\
	W_{i,j}(p) \ket{ji} &= -\sin(\theta) \ket{ij} + \cos(\theta) \ket{ji},
	\end{aligned}
\end{equation} 
where the amplitude $\sqrt{p}$ has been replaced by a periodic function $\cos(\theta)$ for ease of optimization.
The protocol proceeds as follows:
\begin{enumerate}
	\item Apply the $n$-qudit circuit $\hat{V}(\vec{\theta})^n_L$ and maximize the probability of finding the $j=1$ qudit in the state $\ket{n-1}_1$
	\item Apply the $n-1$-qudit circuit $\hat{V}(\vec{\theta})^{n-1}_L$ spanning qudits $[2,n]$ and maximize the probability of finding the $j=2$ qudit in the state $\ket{n-2}_2$.
	\item Repeat this protocol for $j\in[3,n-1]$.
\end{enumerate}
Numerically, we successfully implemented the VQU ansatz validation protocol for up to $n=4$ qudits.

\section{Optical VQU} 
\label{sec:scaling_calculations}

In this section it will be convenient to introduce notation for the probability of seeing between 1 and $k$ photons in the $j^\text{th}$ optical mode, which for a given output state $\ket{\psi_\text{out}}$ is
\begin{equation*}
	\tilde{P}^k_j=\sum_{i=1}^k |\braket{i_j|\psi_\text{out}}|^2.
\end{equation*}

\begin{figure*}[t!]
\includegraphics[trim=0 0 0 0, clip, width=0.9\linewidth]{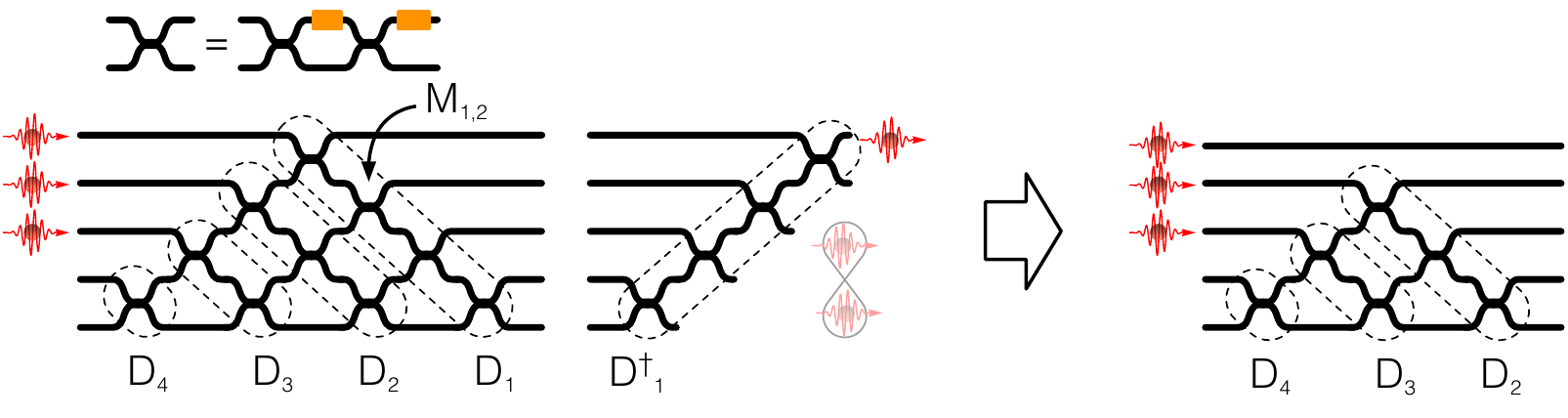}
\caption{\textbf{Compressed Optical Unsampling.} 
A single diagonal circuit $D_1^\dagger$ is sufficient to unentangle a single photon from the first mode.
.}
\label{fig:fig2_app}
\end{figure*}

\subsection{Scaling Calculation} 
\label{ssub:scaling_measurement}

In the following we present calculations for the scaling arguments used in the body of the paper.
First note that the number of ways of placing $n$ in $m$ optical modes (which naturally includes collision events) is given by the multichoose formula

\begin{equation}
	\left ( \binom{m}{n}\right) \equiv \binom{m+n-1}{n}.
\end{equation}

Next, note that for $n$ photons in $n^2$ modes, $\tilde{P}_j^1$ the probability of one and only one photon in a mode goes as one minus the the ratio of the number ways of placing $n-1$ photons in $n^2-1$ modes
\begin{equation}
	a = \binom{(n-1)+(n^2-1)-1}{n-1}
\end{equation}
to the total space
\begin{equation}
	b = \binom{n+n^2-1}{n}.
\end{equation}
which is 
\begin{equation*}
	1-a/b=n/(n^2+n-1)
\end{equation*}
scaling as $\mathcal{O}(1/n)$.  

\subsection{Circuit Scaling} 
\label{sub:circuit_construction}
Now we consider the number of photonic circuit elements for the VQU protocol.  For the case of $n$ photons in $m=n^2$ modes, each layer requires $n^2(n^2-1)$ phase shifters, which because there are $n$ layers in total, requires $\mathcal{O}(n^5)$ total elements.
Note however from the construction of Reck et al.\cite{Reck:1994dz}, that each unitary operation can be written as a product of diagonal matrices $\hat{D}_k$
\begin{equation*}
	\hat{U}= \hat{D}_1 . \hat{D}_2 . \dots \hat{D}_{m-1}
\end{equation*}
where each diagonal 
\begin{equation*}
	\hat{D}_k=\prod_{j=k}^{m-1} \hat{M}_{k,j}
\end{equation*}
is an array of $m-k$ Mach-Zhender interferometers (MZIs)
\begin{equation*}
	\hat{M}_{k,j} = 
	\begin{bmatrix} e^{i\phi} \sin(\alpha/2) & \cos(\alpha/2) \\
					e^{i\phi} \cos(\alpha/2) & -\sin(\alpha/2)
	\end{bmatrix}_{k,k+1}
\end{equation*}
parameterized by an internal phase shift $\alpha$ and external phase shift $\phi$. The index $(k,j)$ labels the (diagonal, mode) the MZI operates on as shown in Fig.~\ref{fig:fig2_app}.
Note, each diagonal acts on at most $m-k+1$ modes, so we can append an inverse diagonal to the system $\hat{D}_1^\dagger$ which undoes the effect of the first diagonal
\begin{equation*}
	\hat{D}_1^\dagger.\hat{U} = \hat{D}_2.\hat{D}_3. \dots. \hat{D}_{m-1}.
\end{equation*}
The resulting unitary therefore leaves the first mode untouched, so an input state of $\ket{1_1 1_2 \dots 1_n}$ will transform with unit probability to 
\begin{equation*}
	\hat{D}_1^\dagger.\hat{U} \ket{1_1 1_2 \dots 1_n} = \ket{1}_1\otimes \ket{\psi}_{2\rightarrow n^2}
\end{equation*}
where $\ket{\psi}$ is some entangled state across modes $2$ to $n^2$.  
Each layer minimally requires a single `diagonal' circuit, comprising at most $2n^2$ elements.  Hence the circuit elements required for the full protocol can be reduced to $\mathcal{O}(n^3)$.
However in some instances over parameterization may in fact accelerate optimization.

\section{Optical VQU using Bucket Detectors} 
\label{sec:squeezing_photons_into_modes}

In this section we show that $n$ photons across $n^2$ modes can be unsampled using so called `bucket detectors' which distinguish between $n=0$ and $n>0$ photon number.

\begin{figure*}[t!]
\includegraphics[trim=0 0 0 0, clip, width=0.9\linewidth]{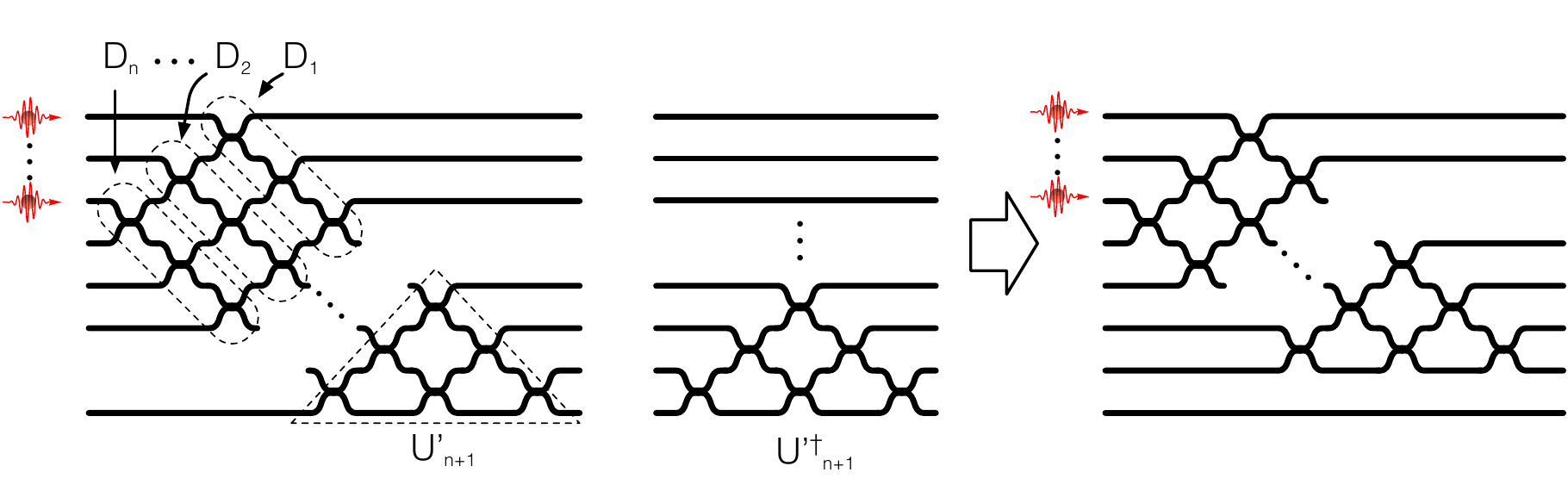}
\caption{\textbf{Squeezing photons into modes.} A single $n\times n$ circuit is sufficient to bring all photons into the first $n^2-1$ modes.  This process is then repeated till all $n$ photons are in the first $n$ modes.}
\label{fig:fig3_app}
\end{figure*}

To see this we must first show that $n$ photons across $n^2$ modes can be squeezed into $n$ modes.
Given the decomposition of Reck et al. an arbitrary unitary operator $\hat{U}$ for $n$ photons in $n^2$ modes can be be decomposed as a product of diagonals followed by a $(n+1) \times (n+1)$ unitary $\hat{U}'_{n+1}$ starting on mode $(n^2-n+1)$ --- see Fig.~\ref{fig:fig3_app}.  Formally we write
\begin{equation}
\label{eq:1}
	\hat{U} = \hat{U}'_{n+1}. \prod_{k=1}^{n} \hat{D}_k
\end{equation}
where each diagonal now stretches $n^2-n$ modes
\begin{equation*}
	\hat{D}_k = \prod_{j=k}^{n^2-n} \hat{M}_{k,j}.
\end{equation*}

Given \eqref{eq:1} it is clear that appending a $\hat{U}'^\dagger_{n+1}$ to the system will result in
\begin{equation*}
	\hat{U}'^\dagger_{n+1} . \hat{U} = \prod_{k=1}^{n} \hat{D}_k
\end{equation*}
which is a unitary that only only spans the modes $1\rightarrow (n^2-1)$
The total result of both $\hat{U}$ and $\hat{U}'^\dagger$ is
\begin{equation*}
	\hat{U}.\hat{U}'^\dagger_{n+1}\ket{1_1 1_2 \dots 1_n} =  \ket{\psi}_{1\rightarrow (n^2-1)} \otimes \ket{0}_n
\end{equation*} 
where $\ket{\psi}_{1\rightarrow (n^2-1)}$ is some entangled state across modes $1\rightarrow (n^2-1)$.  This shows there exists at least one circuit setting whereby getting the state $\ket{0}_{n^2}$ is maximized.
To find this setting one appends a circuit $\hat{U}_{n+1}(\vec{\phi})$ and varies $\vec{\phi}$ to minimize the probability of any event in mode $n^2$.
This minimization of the photon flux can readily be achieved with bucket detectors.
The protocol is then repeated $n^2$ times until all photons are found in the first $n$ modes.
For this protocol, each optimization stage requires $\mathcal{O}(n^2)$ parameters, and the total protocol therefore requires $\mathcal{O}(n^4)$ circuit elements.

Now that all photons are in the first $n$ modes, we show how to recover the initialization state $\ket{\psi}_\text{in}=\ket{1}^{\otimes n}$.
To understand this first note that given one photon input per mode, linear optical operations preserve average photon number per mode.
Let $U$ be an $n\times n$ unitary matrix with entries $u_{i,j}$, and $\ket{\psi}=U \ket{1}^{\otimes n}$ be the state generated by applying $U$ to an $n$-mode Fock state consisting of single photons in each mode.  We wish to compute the average photon number $\bar{n}_i=\braket{\psi|\hat{n}_i|\psi}=\braket{\psi|b_{i}^\dagger b_i|\psi}$ in any given output mode $i$.  If we denote the creation operators of the input modes as $a_j^\dagger$, then $b_i^\dagger=\sum_j u_{i,j} a_j^\dagger$ and it follows:
\begin{eqnarray*}
\bar{n}_i&=&\braket{\psi|b_{i}^\dagger b_i|\psi} \\
&=&\sum_{j,k}u_{i,j}u^*_{i,k}\bra{1}^{\otimes j}a_{j}^\dagger a_k \ket{1}^{\otimes k} \\
&=&\sum_{k}|u_{i,k}|^2\bra{1}^{\otimes k}a_{k}^\dagger a_k \ket{1}^{\otimes k} \\
&=&\sum_{k}|u_{i,k}|^2 \\
&=& 1
\end{eqnarray*}
where the final equality follows from the unitarity of $U$.

We see that a bucket detector in a mode will therefore maximize its on/off count ratio if and only if the output state in that mode is identically $\ket{1}$.  This is because any two or higher photon-number contribution will necessarily be averaged out by the frequency of detecting vacuum.
The protocol proceeds by sequentially maximizing $\tilde{P}^n_j$ for $n\in[1,n-1]$ via a diagonal $2n$ parameter circuit.

\section{Optical VQU Numerical Experiments} 
\label{sec:optical_vqu_numerics}
In the following we describe the optical VQU numerical experiments, for up to $n=6$ photons, presented in Fig.~\ref{fig:fig4} of the main manuscript.
While optimizing $\tilde{P}_j^1$ for $j\in[1,n]$ is sufficient to perform the VQU protocol, in many cases this performs poorly due to (1) the number of parameters involved in the optimization and (2) the absence for an analytic expression of the gradient.
We determined that reducing the parameter set by first bring all photons into the first $n$ photons was superior in terms of speed and accuracy of the unsampling protocol.

To compress $n$ photons into the first $n$ modes we perform the following protocol:
\begin{enumerate}
	\item Generate an $n^2$-dimensional sampling unitary via the Haar measure
	\item Pass the output state into $n^2$-dimensional unsampling circuit with all phases $(\alpha, \phi)_{j,k}=(0,0)$
	\item For $j\in[1,n]$ and $k\in[n^2,1]$ optimize $(\alpha,\phi)_{j,k}$ to minimize the photon flux in the $k+1$ optical mode.
\end{enumerate}

A single iteration of this protocol is sufficient to yield a $>0.99$ probability of all photons in the first $n$ modes.  To achieve numerical accuracy we repeat this three times.

Next, to unsample $n$ photons in $n$ modes we perform the following protocol:

\begin{enumerate}
	\item Pass the output state into a $n$-dimensional unsampling circuit with all phases $(\alpha, \phi)_{j,k}=(0,0)$
	\item Maximize the probability of any event, $\tilde{P}^n_1$, in the $j=1$ optical mode over all parameters $(\alpha, \phi)_{j,k}$
	\item Append a $(n-1)$-dimensional unsampling circuit acting on modes $[2,n]$ and maximize $\tilde{P}^n_2$
	\item Repeat for a $(n-j)$-dimensional circuit and  $\tilde{P}^n_j$ for $j\in[3,n-1]$
\end{enumerate}

In case numerically accuracy is not achieved we allow for random restarts, which is included in the total number of iterations plotted in Fig.~\ref{fig:fig4}.
To estimate the scaling we fit the number of iterations required to reach numerical accuracy, against the photon number, for a range of hypothesis models.  In Table~\ref{tab:VQU_fit_error} we plot the $1-R^2$ error for each model.

\begin{table}[]
\begin{tabular}{l|l}
Model & $1-R^2$ error      \\ \hline
$a+bx$   & $0.12$             \\
$a+bx + cx^2$ & $1.5\times10^{-3}$ \\
$a+bx+cx^2+dx^3$ & $9.6\times10^{-7}$ \\
$a+be^{cx+d}$ & $1.1\times10^{-3}$
\end{tabular}
\caption{Error in fit parameters for optical VQU scaling.}
\label{tab:VQU_fit_error}
\end{table}

\end{document}